\documentclass[12pt,english]{article}
\usepackage[T1]{fontenc}
\usepackage[utf8]{inputenc}
\usepackage{lmodern}
\usepackage{natbib}
\usepackage{csquotes}
\usepackage{setspace}
\usepackage{graphicx}
\usepackage{mathrsfs}
\usepackage{amsmath}
\usepackage{amssymb}
\usepackage{babel}
\usepackage{enumerate}
\usepackage[text={16.5cm,23cm},centering]{geometry}
\expandafter\def\expandafter\quote\expandafter{\quote\small}

\makeatletter

\@ifundefined{date}{}{\date{}}
\makeatother

\begin{document}

\title{\textbf{Special theory of regularity}}

\author{Juliano C. S. Neves%
\thanks{juliano.c.s.neves@gmail.com%
}}

\maketitle

\begin{center}
{\it{Instituto de Ciência e Tecnologia, Universidade Federal de Alfenas, \\ Rodovia José Aurélio Vilela,
11999, CEP 37715-400 Poços de Caldas, MG, Brazil}}
\end{center}

\vspace{0.5cm}

\begin{abstract}
The theory of regularity is a philosophical perspective in which laws of nature are just descriptions, that is to say,
laws of nature do not govern the world. Moreover, according to the theory of regularity, the number of laws of nature
might be infinite, thus any attempt towards the theory of everything is doomed.
 Here I propose a special or restricted theory of 
regularity. The main difference as to the well-known version of that theory is both the range of validity and the 
scale of the laws of nature. Laws of nature ought to be considered just inside the observable universe 
and within certain energy and length scales. 
Even so I apply the theory of regularity to the multiverse scenario. 
As a consequence, the special theory of regularity supports
only two types of multiverses by comparison with our world: those ones with a different sequence of unique events 
and different laws of nature and those ones with the same sequence of unique events and the same 
laws of nature instanced by the unique events. 
The latter case is some sort of eternal recurrence or a parallel eternal recurrence.       
\end{abstract}


\section{Introduction}
The answer for the question \enquote{what is a law of nature?} is not as simple as it seems to be. 
The concept of law of nature is mainly a modern concept. 
Neither Plato nor Aristotle, for example, thought of laws of nature. Instead of laws,
the two great ancient philosophers had thought of metaphysical principles on which our world was grounded. 
For \citet[48a]{Plato}, in particular, there were two opposite principles that gave rise to the physical world, namely
\textit{nous} and \textit{ananke}. 
The former is the principle of order, and the latter is the principle of disorder. In the Platonic cosmogony and cosmology,
 described in the \textit{Timaeus}, \textit{nous} is also  
the intellect attributed to the God-Demiurge, the Creator of the cosmos,\footnote{For details and 
comments on Plato's cosmology, see \cite{Brisson}, \cite{Carone}, \cite{,Neves3,Neves4}, and \cite{Zeyl}.} 
and the principle of \textit{ananke} is interestingly translated into English as necessity. 

But in modernity the concept of law of nature emerges as something that should either describe or govern the world.
In this regard, philosophy of science presents a hard battle between regularists and necessitarians.\footnote{Some 
authors try to suggest alternatives to that dichotomy (see, for example, \citealt{Bird}).}   
Members of the regularity party say that the theory of regularity avoids 
either the ancient belief in the Demiurge  or the modern belief in metaphysical principles that govern
the world, thus the theory of regularity would be the 
most appropriate philosophical perspective on laws of nature
in agreement with a naturalistic philosophy of science or a world view without (an excess of) metaphysical principles. 
On the other hand, necessitarianism appeals to a modal feature, given by the nomological necessity, such that laws of 
nature are necessary principles on which the world is grounded.\footnote{See, for example, \cite{Armstrong}.}
 The old belief in something true beyond the empirical world
is still considered in the necessitarianism.  For regularists, laws of nature are descriptions\footnote{Throughout 
this article, description is 
synonymous with interpretation from Nietzsche's philosophy. In Nietzsche, interpretation means that
what is interpreted does not reveal something beyond the phenomena. That is, \enquote{inasmuch as the word 
\enquote{knowledge} has any meaning, the world is knowable: but it is variously \textit{interpretable}. It has no meaning
behind it, but countless meanings} (\citealt[p. 139, fragment 7(60)]{Nietzsche2}). See \cite{Neves1} for
a brief discussion on the notion of interpretation and laws of nature in Nietzsche.} (regardless of whether the description
comes from experiments or \textit{Gedankenexperimente}), but for necessitarians
laws of nature govern the entire world.  According to \citet[p. 38]{Swartz}, for a regularist the laws of nature
\enquote{derive their truth from the actual (i.e., instanced) connections
(between states and between events) in the world.} Therefore, the laws of nature \enquote{express only 
what \textit{does} occur.} On the other hand, for a necessitarian the laws of nature 
\enquote{determine which connections can and cannot occur (...) express what \textit{must} 
occur in particular circumstances.} Thus, the modal element is present only on the necessitarian side of this debate. 

It is worth emphasizing that it is possible to think about laws of nature as relations among universals \citep{Armstrong},
universal properties \citep{Dretske},
disposition of properties \citep{Bird}, and laws of nature as something inserted into \enquote{integrated systems} (\citealt[p. 367]{Lewis}) even in the theory of regularity perspective. Lewis' \textit{best systems account} is
a very popular alternative in the theory of regularity, however, as we will see, I adopt here Swartz's alternative \citep{Swartz}
because his discussion on unique events is more appropriate to the multiverse scenario and its relation to laws
of nature. 
But regardless the regularity theory version, the dichotomy theory of
regularity--necessitarianism is brought out by the very difference (for necessitarians) or not (for
regularists) between contingently true universal propositions and laws of nature. For regularists, there is no difference 
whatsoever, however for  necessitarians there would be an important difference, because only laws of nature would 
express the nomological necessity \citep{Armstrong,Carroll}.

As I said, the perspective adopted by me in this debate between regularists and necessitarians is the theory of regularity. 
But here I propose a modification of the theory.  Laws of nature from a reasonable theory of regularity---considering its 
very \textit{essence}---should be limited to the observable universe, which is a concept derived from 
the invariance and limit of the speed of light in vacuum in the theory of relativity. Beyond the cosmological horizon, 
neither laws nor conclusions should be drawn scientifically. As laws of nature are descriptions of phenomena, 
according to the theory of regularity, thus 
it is reasonable to look for descriptions just inside the observable universe, which is the set of all
phenomena. This part of discussion is based on the degree of scientificity proposed recently in \cite{Neves2}. According to
that degree, in which fuzzy sets were adopted in order to quantify a notion of scientificity, 
without data or phenomena the degree of scientificity of any
statement is zero. Therefore, laws of nature from some scientific evidence or \textit{fact} should work just inside 
the observable universe.

After restricting the laws of nature to the observable universe, a version of 
the theory of regularity, which I call \textit{special theory of regularity},
 is applied to the multiverse scenario. 
But before that, the state-description model is presented. 
The state-description model developed by \cite{Carnap} and applied to the theory of regularity concerns 
the reduction of universal (or statistical) propositions
to singular \textit{facts}, which are translated into singular propositions. 
From the state-description model, the set of universal propositions of possible
worlds and sequences of unique events (or events that are one of a kind) are studied, and the conclusion
from the theory of regularity is given: \textit{given two possible worlds
with different sets of unrestricted universal propositions, thus such worlds have different laws of nature}.
  
Then with the aid of the state-description model applied to the theory of regularity, the multiverse scenario is
studied in such a philosophical context. 
 The multiverse perspective is the very speculation according to which our universe is not just one, 
but there would exist a huge number of bubble universes (also called parallel universes).
 According to \cite{Tegmark}, there would be types of
multiverses or levels of multiverses, some of them with either identical or different laws of nature
when compared to ours. Using the special theory of regularity, 
I argue that there could exist just two types of worlds in the multiverse scenario (compared to our world): 
those ones with both identical sequence of unique events and identical laws of nature that were instanced during
the sequence of unique events, and 
 those ones with both different sequence
of unique events and, at the same time, different laws of nature. There could be just two alternatives 
within this perspective. 
There would not be worlds in which the outcome of a unique event would be different from our world and the 
laws of nature would be the same. This possibility is ruled out according to the theory of regularity presented here.
From that conclusion, a scenario of infinite worlds would provide a type of parallel eternal recurrence, in which 
worlds with exact outcomes of unique projects (and with the same laws of nature that were instanced by
the sequence of unique events) would coexist in different 
phases or proper times.

The article is structured as follows: Section 2 is about the meaning of a law of nature and the conditions for
lawfulness. A short introduction to the theory of regularity is presented in Section 3. A special or restricted version of the
theory is suggested in Section 4. In Section 5,
 the restricted theory is applied to the multiverse scenario addressing the types of alternative or parallel worlds.
 The final considerations are given in Section 6.

\section{What is a law of nature?}
\label{What is a law of nature}
Following \cite{Swartz}, necessary conditions for physical lawfulness are indicated in this section.\footnote{For
an in-depth review on laws of nature see \cite{Carroll}.} 
In general, scientists do not have a sophisticated 
definition of  law of nature or even clear conditions for lawfulness.  
In this regard,  philosophy of science could help them. 
It is common in philosophy of science to indicate some  (necessary and maybe not sufficient) features or
conditions of a proposition that is candidate for law of nature. Accordingly, laws of nature 
are (i) \textit{factual} truths, not 
logical ones, (ii) are true for every time and place in the universe (in agreement with the cosmological principle), 
(iii) do not contain proper names, (iv) are universal or statistical statements, 
(v) and are conditional statements, not categorical ones. These conditions, according to
\cite{Swartz}, are necessary ones. Moreover, they are sufficient
conditions for the theory of regularity and not for necessitarianism, because necessitarianism
 needs an extra condition: the nomological necessity. Before commenting the extra condition (for necessitarians),
 let us see in details the five primary conditions for lawfulness, which will be modified later in the cosmological
 context:

\begin{enumerate}[(i)]

\item The first condition, which says that laws of nature are \textit{factual} truths,\footnote{See \cite{Cartwright}
for a criticism on this point. Cartwright says laws of nature, constructed by composition of causes, fail in
stating true facts.} not logical truths, 
means that physical lawfulness concerns the empirical world, or physical world as \citet{Swartz} says. Laws of 
nature are true from the phenomena, not from a priori or logical propositions. Moreover, a candidate for 
law of nature is not necessarily true from the logical point of view (like two plus two equals four), it is 
logically contingent, that is to say, such a proposition could be different in our world. 
For example, the inverse-square law (hypothetically considering it as a candidate for law of nature) 
could be different, it is not logically necessary.  

\item The second condition or feature says that laws of nature are true for every time and place in the universe. 
As I mentioned
before, this feature is in agreement with the cosmological principle, the main principle in cosmology today. 
The cosmological principle states that every freely falling observer is equivalent in the universe. As a consequence,
the universe for that class of observers is homogeneous and isotropic on large scales. This is a good
point in which I would modify the theory of regularity (concerning where these observers are equivalent).
 The  point to be corrected is the range of validity of the laws. 
 As the theory of regularity is a perspective based on the phenomena, laws of nature would be
  more appropriate inside the observable universe (to be defined in section \ref{Constraining}).

\item A law of nature does not contain proper names, just general descriptions. It is not about this $A$ and that $B$, 
it is about $A$ and $B$ in general, with $A$ and $B$ playing the role of general properties, not particular properties
 like the mass of Sun (a law of nature speaks of masses of stars of the same category of the Sun). The law of inertia says
 about bodies in general, not just about my or your body. 

\item Laws of nature concern universal or statistical propositions, that is to say, given the object 
$O$ and its property $A$, a law of nature speaks of all $O$s and their properties $A$s (universal
case) or of a fraction of $O$s with the property $A$ (statistical case). For example, the radioactive decay
instances a statistical law of nature. 

\item A law of nature is written as a conditional statement or proposition, not as categorical one. 
Conditional propositions like
\enquote{If $A$, then $B$} can be laws of nature. On the other hand, a categorical statement like
 \enquote{$O$ is $A$} is not a good candidate for lawfulness. But it is possible to write some categorical
 statements as conditional ones, even laws written by means of equations are able to be written as conditional
 statements. For example, the law of inertia would be like this one in the conditional form: if no resultant 
 force is acting upon a body, then such a body will be at rest or in uniform motion.  
 
\end{enumerate}

According to \cite{Swartz}, the extra ingredient for necessitarianism---the nomological necessity---is not  
equal to physical necessity. The latter says that given the set of laws of nature \textbf{L}, 
the proposition $p$
in agreement with \textbf{L} follows necessarily from \textbf{L}. That is,
\begin{equation}
\boxdot p = \textbf{L} \rightarrow p, 
\label{PN}
\end{equation}
in which $\boxdot$ means \enquote{it is physically necessary that}, 
and the operator $\rightarrow$ stands for logical implication
(something stronger than the material implication).
 Thus, it is physically necessary that $p$ is implied by \textbf{L}. 
The physical necessity is not a condition for lawfulness because it assumes the set of laws of nature \textbf{L}. If 
the physical necessity was
a condition among the five conditions, it would be a circular reasoning made by us. 
Because of that, \citet[p. 46]{Swartz} calls
physical necessity \enquote{a weaker, or degenerate, kind of necessity (...).} 

However, for necessitarians, the nomological necessity would be something more. As I said, the theory of regularity needs 
no additional principle or nomological necessity in order
to claim a law of nature. On the other hand, according to necessitarianism, 
the notion of nomological necessity expresses the necessity
of law, that is to say, it is a hypothetical principle added in order to guarantee that a law of nature is something different
from contingently true propositions. As mentioned by Swartz, the main problem is that the nomological necessity
is a metaphysical notion, something that is not provided by the empirical or physical world, it is 
some sort of heritage from the deistic world view (\citealt[p. 202]{Swartz}). I dare to say that the nomological
necessity is some sort of shadow of the dead God denounced by Nietzsche, that is,
such a shadow is a metaphysical truth created (like God) in order to justify the world in which we live.\footnote{According to the  
aphorism 108 from the book \textit{Gay science}, Nietzsche says:: \enquote{God is
dead; but given the nature of people, there may still be for millennia caves wherein
they show his shadow.---And we---we must still defeat his shadow as well!} Also, 
 the big bang was interpreted as God’s shadow by \cite{Neves1}.} 

Above all, the nomological necessity has two main functions: to separate contingently true propositions from nomological
truths, which are laws of nature for necessitarians, and to explain or justify 
the world regularity by means of a metaphysical
principle. The cosmos or order in Greek would come from and be maintained orderly by the nomological necessity. 
For necessitarians,  a large number of true propositions that satisfy the five conditions above is not
eligible for lawfulness, because those true propositions are contingent. 
The set of that huge number of true propositions would be larger than the set of nomological truths.
Therefore, according to necessitarianism, the number of laws of nature is very small.
 On the contrary, the theory of regularity (in 
Swartz's version) accepts even an infinite number of laws.

I would like to mention an influential and popular (among scientists) notion of laws of nature given by 
\citet[p. 19]{Popper}, according
to which \enquote{the more they [the laws of nature] prohibit the more they say}, or \enquote{natural laws 
\textit{forbid} certain events to happen (...) they have the character of \textit{prohibitions}} (\citealt[p. 449]{Popper}). 
Popper was a necessitarian,\footnote{Even considered by many a regularist or an imperfect regularist according to
 \citet[p. 107]{Swartz},
there are scholars who defend Hume as a necessitarian, when he speaks of \enquote{necessary connexion
betwixt causes and effects} (\citealt[p. 165]{Hume}). See \citet[p. 39]{Swartz} and references therein.} 
and this point is made clear from the following quotation:
\begin{quote}
I believe (...) the idea that there are necessary laws of nature, in the sense of natural or
physical necessity (...) is metaphysically or ontologically important, and of great intuitive
significance in connection with our attempts to understand the world. And although it is
impossible to establish this metaphysical idea either on empirical grounds or on other grounds, 
I believe that it is true (...) (\citealt[pp. 459-460]{Popper}).
\end{quote} 
As we can see, the necessity that Popper calls (here called nomological necessity) is neither falsifiable nor (even worse)
something able to be established on empirical grounds. Even science today working upon an unthinkable amount of data,
 scientists still share a metaphysical view on laws of nature. 
The only one alternative to this is the theory of regularity.

\section{Brief introduction to the theory of regularity}
\label{Theory}
As mentioned before,  the theory of regularity and the necessitarianism assume implicitly or not
the five conditions for lawfulness described in Section \ref{What is a law of nature}, but
 necessitarianism also needs the nomological necessity in order to \enquote{explain}
the laws of nature. Because of that, necessitarianism 
splits true universal propositions that satisfy that all five conditions into nomologically necessary propositions
 and contingent propositions. Only nomologically necessary propositions would be laws of nature. 
On the contrary, for the theory of regularity, a statement
that satisfies that five features is contingent and, at the same time, is a law of nature.
Therefore, there may be an infinite number of laws of nature for regularists.  As \citet[p. 101]{Swartz} says:
\begin{quote}
There are a number (an infinite number, perhaps) of true universal, material conditional
statements (propositions), all of whose terms are perfectly descriptive, that is, make no reference
to any particular time, place, person, or thing in the world. The Regularist is content to allow that
all these true universal, material conditionals are physical laws, but the Necessitarian wishes to
divide this class into two mutually exclusive and jointly exhaustive subclasses: the
nomologically true and the accidentally true. 
\end{quote}
Then an interesting question as to the theory of regularity concerns the number of laws of nature or
the infinite number of laws of nature. 
In this point, I would make clear the terminology adopted here. \citet[p. 3]{Swartz} claims that there is the very difference
between scientific laws and physical laws. The former means approximations adopted in the daily work by scientists, 
and the latter are universal and general truths that satisfy the mentioned five conditions.\footnote{Here I point out a 
mistake made by \citet[p. 19]{Swartz}, where he said that Kepler's laws are not physical 
laws (or laws of nature as I adopt in this article), but scientific laws.
The main reason for Swartz would be the fact that Kepler's laws concern just the solar system. But after Swartz writing his
book, Keplerian motion was observed in several systems outside the solar system. See, for example,
 \cite{Eisenhauer}, where stars in Keplerian motion were studied around Sagittarius A*.} 
The meaning that Swartz attributes to physical laws is about the same 
that it is adopted in this article as to laws of nature. Then having made clear this point, the number of laws of nature
(or physical laws for Swartz), 
according to theory of regularity, is not small, because new type of phenomena might be described by using new laws. 
A new phenomenon, like the recent jets observed in massive black holes,\footnote{Jets observed, for 
example, in the first image of a black hole composed by the Event Horizon Telescope Collaboration \citep{EHT}.} 
could be described from known and
brand new laws. As the entire world always amaze us with new phenomena, which fall or not under known laws,
new descriptions or new laws are ever required. 

In this regard, according to the theory of regularity, the large number of laws of nature 
would discourage any attempt to reduce the laws of nature to a very small or, even worse, a unique set of laws. 
The sought-after unification of all fields in physics or the
\enquote{theory of everything}, in which the set of 
laws of nature would be reduced to a few laws, inserted into a unique and general theory, is something doomed. 
First of all, it is doomed
 because the number of laws of nature is huge (maybe infinite). Secondly, we will never be able to conclude at any time 
 whether or not we know all laws, that is, we will not be able to answer what everything is. 
 A bit of history of science can convince us about that: for example, the 
 black-body radiation, a new phenomenon in the end of the nineteenth century, gave rise to new physics, 
 the quantum physics \citep{Planck}. Then 
 new laws might be instanced in the future, even after the sought-after unification in physics. 
 The unification of fields in physics
 has had successful examples, like the electromagnetic field or the electroweak interaction \citep{Glashow,Salam,Weinberg1},
  which is the unification of both the electromagnetic field and the weak field (responsible for the 
  radioactive decay of atoms).\footnote{However, the unification of quantum mechanics and 
 general relativity is still in progress.}
But reducing the number of laws to a small number, disregarding new future phenomena
 in which new laws would be instanced, is matter of faith or one more shadow of the dead God. 
 Even so, the search for unifying phenomena is valuable as the electromagnetic theory showed us more than a century
 ago. But, according to theory of regularity, the \enquote{theory of everything} is impossible.       

Another important point in the theory of regularity concerns physical possibility. What is physically possible 
or impossible?
According to the theory of regularity and necessitarianism, 
that which is congruent with all laws of nature is physically possible.
And that which is not congruent with all laws of nature is physically impossible. This point is shared
by either perspectives. However, there is a notable divergence as to the physical impossibility. 
For regularists, something is physically impossible just because it did not/does not/will not exist, then its existence is a
 timeless falsity, with no example of it. It is just a \textit{factual} argumentation. But for necessitarians, 
 there is a modal element in this problem, for to say something is physically impossible
  means that it \textit{could not} exist at 
any time. And, of course, for the necessitarian view the laws of nature tell us 
 what \textit{could not} exist.
 
The theory of regularity assumes the state-description model, in which laws of 
nature (true universal propositions) are reduced to singular propositions that describe the state of affairs of 
the world. \citet[p. 81]{Swartz} writes:
\begin{quote}
(...) we may say that a \textit{state-description} of a possible world $W$ is the class of (or the conjunction of)
 all the true propositions 
obtaining in $W$. Then, in terms of this extended concept of a state-description, the Regularity Theory may be
expressed thus: All physical laws of a possible world $W$ are reducible to some subset
(simplification) of the state-description of $W$. In other words, Regularists disallow that there are
any actual or possible physical laws whose truth derives from anything other than a world's
\textit{singular} facts.
\end{quote}
The main problem of the state-description model is answering this question: how are true general propositions 
 reduced to conjunctions of singular propositions? For instance, given the set $\textbf{S}$ of $n$ propositions in the
 form \enquote{if $A$, then $B$}:
\begin{align}
\forall x_1& (A x_1 \supset B x_1), \nonumber \\
\forall x_2 & (A x_2 \supset B x_2),  \nonumber \\
& \vdots  \nonumber \\
\forall x_n & (A x_n \supset B x_n), 
\label{S}
\end{align}
where the symbol $\forall x$ stands for \enquote{for any $x$}, $n$ is finite and integer positive  
(in this example, $x_n$ is about $n$
elements), and the operator $\supset$ indicates material implication. 
The universal proposition $U$, also called packing relation, that would generalize \textbf{S}, indicated by  (\ref{S}), is 
written as
\begin{equation}
\forall x(Ax \supset Bx). 
\label{U}
\end{equation}  
But we only reduce $U$ to \textbf{S} or (\ref{U}) to (\ref{S}) with the aid of an auxiliary premise $P$, given by
\begin{equation}
\forall x (x=x_1\vee x=x_2 \  ... \vee x=x_n ),
\label{P}
\end{equation}
where $\vee$ is the logical disjunction, which stands for \enquote{or}. 
However, the premise $P$, indicated in Eq. (\ref{P}), is itself  universal.
 Therefore, in order to reduce a universal proposition to singular propositions we needed another
universal proposition. 
Then, it would be, in principle, impossible to reduce a universal proposition to a set of singular propositions
or, equivalently, it would be impossible to pack \textbf{S} into $U$. 
But \citet[p. 87]{Swartz} argues that the mentioned problem is just because one concludes from an unspecified
world. From a specific world, in which $P$ is valid, $ U \leftrightarrow$ \textbf{S} 
would be logically false, but
\begin{equation}
\square \left[\forall x ( x=x_1\vee x=x_2 \  ... \vee x=x_n ) \supset (U \equiv \textbf{S}) \right] 
\end{equation}  
would be true.\footnote{Here $\square$ means \enquote{it is necessary that}.} \textbf{S} and $U$ do not form a logical implication (valid in all possible worlds), that is to say, 
\textbf{S} $\rightarrow U$ is false. But  $U \equiv$ \textbf{S} is valid in all possible
worlds in which $P$ is true. Thus, the packing relation of $\textbf{S}$ given by $U$ shows
that $U$ is reducible to $\textbf{S}$. As \citet[pp. 87-88]{Swartz} says: \enquote{The packing relation, 
although not the relation of logical implication, is nonetheless far stronger than
mere material implication, is perfectly intelligible, and is free of internal incoherence.} 

The main result of the theory of regularity that will be adopted in Section \ref{Multiverse} 
is about possible worlds with 
different unique events, i.e., \textit{events that are one of a kind}. Such unique events (or unique projects) 
are interesting because they can provide
true universal propositions (not statistical ones). 
The state-description model says that given two possible worlds $W$ and $W'$ and a unique
event (or a unique project of a kind) that differs either worlds, for instance an event 
that occurs (it is successful) in $W$ but not in $W'$, then the two worlds will have
different laws of nature. For instance, given the unique event $E$ (to run 100 meters in less than 9 seconds)
 and its successful result $R$ (someone ran 100 meters in less than 9 seconds), and just one
instance $x_1$ of $E$, in which it is successful in $W$ but it fails in $W'$. Then, considering all time, 
past, present and future, one has different universal
propositions in each possible world:
\begin{equation}
\forall x_1(E x_1 \supset R x_1) \   \mbox{for $W$},
\label{W}
\end{equation}  
and
\begin{equation}
\forall x_1( Ex_1 \supset   \neg R x_1)  \    \mbox{for $W'$},
\label{W'}
\end{equation}
in which $\neg$ means negation. 
Both worlds would differ from a true universal proposition and then from a law of nature. That is to say:
\begin{quote}
If being a true contingent universal generalization having unrestricted terms is a sufficient condition for being a
physical law, then two possible worlds that differ in the outcomes of their respective unique projects must 
differ in their physical laws \citep[pp. 89-90]{Swartz}.
\end{quote}
Keep in mind that (\ref{W}) and (\ref{W'}) are general propositions and contain no proper names. 
Moreover, the unique event $E$ could be any event that is
unique in either worlds.  As each unique successful event is, according to the theory of regularity, 
something physically possible, something that falls under a set of  laws of nature, the success of such an
event in the world
$W$ and its bad outcome in the world $W'$ are justified by there being different laws of nature in each world.
 The unique event that did not 
happen in $W'$ shows that such an event, to run 100 meters in under 9 seconds, 
was physically impossible in that world. For that example, \enquote{nobody can run 100 meters in under 9 seconds}
 would be a law of nature in $W'$ (not in $W$),
  in agreement with Section \ref{What is a law of nature}, in the conditional form: 
\enquote{if something is human, then he/she can not run 100 meters in less than 9 seconds.}

\section{Restricting the laws of nature}
\label{Constraining}
Today's cosmology is Einsteinian, that is, the recent cosmology is derived from the theory of general relativity or,
from the regularity point of view, data are interpreted (or described) by using a solution of Einstein's field equations 
\citep{Einstein}.
The cosmological spacetime (or cosmological geometry that is a solution of the field equations)
 is the Friedmann-Lemaître-Robertson-Walker spacetime \citep{Friedmann,Lemaitre,Robertson,Walker}, which is
expanding because of the dark energy \citep{Riess,Perlmutter}. Thinking backwards in time, one concludes that
the universe was smaller and hotter in the past. Thus, such dynamical spacetime would suffer from 
the singularity problem or
the big bang problem, which is ill interpreted as beginning of time. But according to
 \cite{Neves2}, singularities and
even the initial singularity or the big bang are nothing but a problematic concept in physics.
Singularities are some sort of noumenon or a problematic
concept in a Kantian reading. A problematic concept is something \enquote{that contains no contradiction
but that is also, as a boundary for given concepts, connected with other cognitions, the objective
reality of which can in no way be cognized} (\citealt[B 310]{Kant}). The big bang or a singularity 
as a problematic concept is some sort of
limit or a \enquote{boundary} for the well-known concept of geodesic in general 
relativity, it has no reality.\footnote{Similar conclusion is drawn in \cite{Romero}.}  

Also the Einsteinian cosmology assumes both the matter-energy content from a fluid description and, as an important
ingredient, the cosmological principle commented before. The Einsteinian cosmology gives rise to the standard model
in cosmology, in which the inflationary mechanism is added in order to generate a universe like ours.\footnote{Developed by \cite{Starobinsky}, the
inflationary mechanism will be discussed in Section \ref{Multiverse}.}

From the postulate (or law of nature) according to which \enquote{nothing travels faster than light in vacuum}, general
relativity provides different types of horizons.\footnote{See \citet[p. 98]{Weinberg2}.} In particular, one has 
the cosmological horizon (or particle horizon) in the cosmological
context, from which the notion of observable universe is derived. The observable universe is a sphere of 
about 46 billion light-years in radius (or $4.4 \times 10^{26}$ meters), whose center is the planet Earth or our position as observers. Because of the 
cosmological principle, an extraterrestrial life living in a distant galaxy (a freely falling observer like us) 
would see the same observable universe.
 The main point here is that beyond the observable universe everything is unknown for us humans.  Someone could
argue that it is possible to speak of phenomena beyond the observable universe because such a speculation is
from reliable and well-tested theories like the theory of general relativity. This is a good point, indeed 
one can speculate it, but for a \enquote{thing} outside the observable universe there will be neither data nor
 any observation.
To say something beyond the observable universe would be \textit{fiction} or mere philosophical speculation.
But here we are talking about laws of nature, that is to say, the theory of regularity is a philosophical
perspective, however laws of nature are subject of science. And there is no science beyond the observable universe, that is,
if there is a degree of scientificity as defined in \cite{Neves2}, no data will mean zero degree of scientificity for 
\enquote{things} outside the observable universe.
As time passes the observable universe grows, because of the universe expansion, and as light reaches us new objects 
and structures get inside the field of view of our most advanced telescopes. 
Then the possibility of new laws describing such new phenomena is not entirely ruled out. 

The mentioned degree of scientificity is built from the Kantian difference between thinking of an object and 
knowing an object. For \citet[B 146]{Kant}, 
physical knowledge needs concepts from the understanding, which is the human
\enquote{faculty for thinking of objects} \citep[B 75]{Kant},
and data from the phenomena. On the other hand, just thinking of something is not knowledge.
 Thus, in \cite{Neves2}, a notion of scientificity is defined
from fuzzy sets, that is to say, it supports the notion of degree.
Given the fuzzy set of theoretical objects (concepts given by the understanding) 
and the fuzzy set of data, knowledge is the intersection of either sets.
In the fuzzy sets theory, the intersection of two sets, in which one of them is empty, 
provides zero degree. For example, theoretically dark energy might be conceived of as the cosmological constant, a
scalar field or even a consequence of spacetime geometry, and moreover there are precise data about the expansion
of the universe \citep{Riess,Perlmutter}.
 Thus, the degree of scientificity of dark energy is not zero. Whether theoretically or empirically,
dark energy presents nonzero degree. But if there were no data as to
the accelerating universe, the degree of scientificity of dark energy would be zero, and the accelerating expansion of the 
universe would be mere speculation. 
By the same argument, the degree of scientificity of models beyond the observable universe is zero, because there are
no data.    

As I support the regularist perspective,  it seems natural to speak of phenomena description inside the observable universe.
Thus, laws of nature that we known from observation and phenomena description are laws of the observable universe.
Therefore, the condition (ii) for lawfulness should be restricted: 
\textit{laws of nature should be true just in the observable universe}.  
That is to say, the set of laws of nature $\textbf{L}$ should be valid within the following interval:
\begin{equation}
\textbf{L}=\lbrace L_1,L_2,...,L_n \rbrace \ \   \mbox{for} \ \  0\leq d \leq r_{obs},
\label{L}
\end{equation}     
in which $d$ is the proper distance of an object (measured from our position),
 and $r_{obs}$ is the radius of the observable universe. A given law of nature is
valid just within that interval. It is worth emphasizing that it is not dependence on $d$, that is, 
the laws of nature do not depend on the proper
distance (at least up to now). Indeed, Eq. (\ref{L}) says that the set \textbf{L} 
works or makes sense just inside that interval,
inside the observable universe.  
 As each law of nature comes from the phenomena description, an attitude in 
agreement with the theory of regularity should judge laws from the observational point of view. Therefore, it works
just inside the observable universe. In this regard, once again, \citet[p. 81]{Swartz} says: 
\enquote{Regularists disallow that there are
any actual or possible physical laws whose truth derives from anything other than a world's
\textit{singular} facts.} Therefore, that which is outside the observable universe is not a \textit{fact}. 

From the history of scientific ideas, the theory of regularity is more convincing than necessitarianism. As is well known,
concepts change over time, that is to say, the description of phenomena is not something fixed or stable.
 For example, gravity in
Newton is a force given by the inverse-square law, but it is spacetime curvature in Einstein's theory
given by the curvature tensor or Riemann tensor. New developments as to the mathematical tools
and even new phenomena \enquote{asked} for  a new description of gravity, as \cite{Einstein} did it. 
Newton's law is of course still valid, but the Einsteinian description works on larger scales than the Newtonian law does.
Gravity conceived of as spacetime curvature brings out a more detailed description, suitable to new phenomena (like 
gravitational lens, spacetime dragging, etc.), which are not available in the Newtonian law. Therefore,
from the five conditions for lawfulness listed in Section \ref{What is a law of nature}, 
the condition (ii) should also be improved in order to accommodate more detailed
descriptions. Then \textit{laws of nature are true for every time and place in the observable universe, 
but within certain scales}. The
Newtonian theory is not valid anymore for massive objects like black holes. In such a scale, Einstein replaces Newton. 
The theory of regularity
and the world view according to which laws of nature are descriptions are not in trouble with the \textit{evolution} of 
concepts.

\section{Application to the multiverse perspective}
\label{Multiverse}

\subsection{Tegmark's multiverse}
For many scholars, the multiverse hypothesis would be a good response to the fine-tuning problem in cosmology 
\citep{Tegmark,Ellis}.
The fine-tuning in cosmology is about the values of parameters in the inflationary mechanism able to provide
 an almost flat, homogeneous and isotropic universe like ours (as it is observed today\footnote{See \cite{PlanckColl}
 for the latest data from the Planck Collaboration. Also, for an overview on types of fine-tuning arguments in physical 
 sciences, see \cite{Adams}.}). 
 The range of values of those 
 parameters that provide a world like we have seen is tiny, then the parameters would be fine-tuned. However, in an 
 ensemble of  worlds, whose number is maybe \textit{infinite},
 the values of the number of e-folds\footnote{The number of e-folds in the inflationary
 mechanism gives the duration of the inflationary process in order to produce a homogeneous and
 isotropic observable universe.} in the inflationary mechanism, for example, would vary, and the value
 obtained by calculations in our world, by using data, is only one among several alternatives or possibilities. 
 There would be nothing special
 in the obtained parameters. In other possible worlds, these parameters would be instanced  differently. 
 A probabilistic argumentation is adopted here. 
 
 It is worth pointing out that the multiverse scenario is not the unique alternative to the fine-tuning problem.
 Bouncing cosmologies have gained room in cosmology, at least among theoretical cosmologists.\footnote{See, for
 example, \cite{Novello}, and \cite{Brandenberger} for reviews on bouncing cosmologies.} 
 Bouncing models try, in turn, to avoid the big bang singularity and, at the same time, the fine-tuning and the multiverse 
 problem of the inflationary mechanism.\footnote{See \cite{Steinhardt} for criticisms as to the inflationary paradigm and
 \cite{Linde2015} for a reply.} Besides, bouncing models offer mechanisms that could replace the 
 inflationary mechanism, 
 like the ekpyrotic mechanism \citep{Lehners}, which is conceived of as a slow contraction phase 
 that would have occurred before
 our expanding phase. But let us assume hypothetically the multiverse scenario in order to see the consequences of the
 theory of regularity in such a speculative perspective.     
 
The multiverse scenario has a strong popular appeal. It appears in books, series, movies, and etc.
 Why? The reason why is the multiverse makes  part of a very common attitude, which I call it the \textit{probabilistic
reasoning}. I call it so because it is a common feature of our world in the twenty-first century. People reason and argue
from probabilistic statements. It is a common procedure today, according to which aspects of everyday life
are measured from statistics and probabilities. The odds of dying in a plane or car crash or of getting sick are
numbers worldwide announced in newspapers and on internet. People are used to the probabilistic reasoning.
That is the reason why multiverse \enquote{makes sense} today. Immersed into a context in which
probabilities are common tools, the multiverse perspective is not a weird perspective for someone who lives in the
twenty-first century. Therefore, without a Creator or a Demiurge for the universe (something ruled out in the modern era), 
the probabilistic reasoning offers an answer, even offering such an 
answer in probabilistic terms. The cosmos---which means order of the world in Greek---is justified by using probabilities in the multiverse picture.

A popular classification of types of multiverses was proposed by
 \cite{Tegmark}. Accordingly, 
there are four levels of multiverses.  Tegmark's four levels are described as follows:
Level I is  an extension of our observable universe, something derived from the inflationary mechanism.
Different initial conditions in the inflationary period would produce different spacetime volumes with different
expanding phases.
 This kind of scenario would require the same laws of nature of our observable universe.
 Different regions (or different \enquote{universes}) could merge, with parts of one getting inside the observable
 region of other. Level II speaks of universes with different physical constants (or maybe different laws of nature).
According to \cite{Tegmark}, this level of the multiverse hierarchy would be promoted by the chaotic
inflation model \citep{Linde1983}. 
With such a model, different symmetry breaking in the early bubble universes would lead
 to different laws of nature for each bubble universe. Even the fine-tuning problem would be solved. 
 With different values for the cosmological parameters in different worlds, the fine-tuned parameters
 that we have seen instance just a possible value among possible values (nothing special in our world).
  The parallel universes in this case would be causally disconnected. 
Level III is about the many-worlds interpretation of quantum mechanics introduced 
by \cite{Everett}. 
In this case, the wave-function description would create new \enquote{worlds} but without new laws. 
Lastly, Level IV is told to be the \enquote{ultimate ensemble}. In this level Tegmark assumes his Platonic view, in which 
mathematical objects would exist. Unknown equations would describe different worlds with very different laws of nature
(some sort of modern art painting). 

Above all, according to the theory of regularity, 
Tegmark's classification leads to unknown worlds. As laws of nature depend on the observation of regularities,
the multiverse levels are some sort of metaphysical speculation, for they speak of 
universes beyond the observable universe.

\subsection{The multiverse perspective in agreement with the theory of regularity}

However, from the theory of regularity briefly described in Section \ref{Theory} and restricted in Section \ref{Constraining}, 
the options to the multiverse hypothesis are
fewer, that is to say, there would be just two types of multiverses:
\begin{itemize}
\item[$\bullet$] Type-1 multiverse: worlds with exactly the same sequence of unique events of our world, thus with the same laws of nature of our world that were instanced during the sequence of unique events.

\item[$\bullet$] Type-2 multiverse: worlds with different sequence of unique events when compared with our world, 
thus with different laws of nature that were instanced during the sequence of unique events.
\end{itemize}
The type-2 multiverse is straightforwardly obtained from the analysis of Section \ref{Constraining}, for
just one different outcome for a unique event in each possible world would lead to different laws of nature in 
each world. On the other hand, for the type-1 multiverse the situation is quite different. As the sequence of 
unique events instances $\textbf{L}^*$, which is the subset of laws of nature described because of that sequence,
 and as the set of laws of nature of our world is $\textbf{L}$, one has
 $\textbf{L}^* \subseteq \textbf{L}$. Thus, the parallel universes with the same sequence of unique events
 share $\textbf{L}^*$, not $\textbf{L}$. Only the set of 
 laws of nature instanced during unique events coincides among these worlds.
  Eventually, there would exist worlds that share $\textbf{L}$ with
 accidental differences, but \textit{maybe} these ones would be more rare. 

A type-2 multiverse would be ever unknown, that is, its laws would be 
absolutely unknown because it would be ever outside the observable universe. Even a type-1 multiverse
would be outside our observable universe, but it would be a copy of our world as for the unique events 
(as defined in Section \ref{Theory}). Thus, for a type-1 multiverse, a kind of eternal recurrence of the 
same unique events could be stated from that sequence of events.
 As is known, \citet[p. 194, \textsection  341]{Nietzsche} speculated about a world in which everything
  returns, \enquote{all in the same succession and sequence (...).} Here, using the theory of regularity,
   the eternal recurrence is not of everything, 
 but just of the unique events of a kind. The conception of the eternal recurrence in Nietzsche is akin to cyclic
cosmologies,\footnote{See \cite{Neves1} for a commentary on the Nietzschean 
eternal recurrence from our latest cosmological 
models.} but here the recurrence is argued in the multiverse scenario, some sort of parallel recurrence, in which 
parallel worlds share the same set of laws of nature instanced by unique events. 
In a scenario in which there exists an infinite number of worlds,  the same sequence of unique events at different times occurs eternally, \textit{in parallel}.        

\section{Final comments}
Taking in consideration our time, the theory of regularity is the most appropriate alternative.
 Without a metaphysical principle or a hypothetical nomological necessity, as the necessitarianism supports it, 
 the theory of regularity is akin to the naturalistic philosophy of science. For the theory of regularity, laws of nature are
 descriptions, are translated into contingent true propositions, and there is no need for an extra ingredient like 
 the nomological necessity promoting lawfulness. As the number of laws of nature is unknown, because
 new instances might bring out new laws in the future, the theory of
 regularity denies any attempt at promoting the theory of everything. 

The special theory of regularity proposed here
restricts the laws of nature to the observable universe in order for the descriptions to be 
considered just from the phenomena.
Moreover, as the gravitational theories of Newton and Einstein indicate, 
laws of nature ought to work on scales of validity,
which are different mass, energy or even length scales.
For massive black holes, the Newtonian law is not valid anymore when considered effects like gravitational lens or
the spacetime dragging generated by the black hole rotation. Here Einstein replaces Newton.

As a consequence of the state-description model adopted in the theory of regularity, two possible worlds that 
differ from  a unique event (one of a kind) could be different as to the set of laws of nature in each world.  
When applied to the multiverse scenario, the special theory of regularity supports just two types of multiverses or
parallel universes when compared to ours: 
(i) those ones with the same sequence of unique events and the same laws of nature instanced by the unique events and
(ii) those ones with different sequence of unique events and different laws of nature instanced by the unique events. 
The first case could be conceived of as some sort of eternal recurrence, a parallel eternal recurrence.

\section*{Acknowledgments}
I thank the Federal University of Alfenas for the kind hospitality.


\begin{thebibliography}{}


\bibitem[Adams(2019)]{Adams}Adams, F. C. (2019). The degree of fine-tuning in our universe --- and others. 
\textit{Physics Reports, 807}, 1-111. https://doi.org/10.1016/j.physrep.2019.02.001

\bibitem[Ade et al.(2016)]{PlanckColl}Ade, P. A. R. Planck Collaboration et al. (2016). Planck 2015 results XIII. Cosmological parameters. \textit{Astronomy and Astrophysics, 594}, A13. https://doi.org/10.1051/0004-6361/201525830

\bibitem[Akiyama et al.(2019)]{EHT}Akiyama, K. Event Horizon Telescope Collaboration et al. (2019).
First M87 event horizon telescope results. I. The shadow of the supermassive black hole.
\textit{The Astrophysical Journal  Letters, 875}, L1. https://doi.org/10.3847/2041-8213/ab0ec7

\bibitem[Armstrong(1983)]{Armstrong}Armstrong, D. M. (1983). \textit{What is a law of nature?}. 
Cambridge: Cambridge University Press.

\bibitem[Bird(2005)]{Bird}Bird, A. (2005). The dispositionalist conception of laws. \textit{Foundations of Science, 10},
 353-370. https://doi.org/10.1007/s10699-004-5259-9

\bibitem[Brandenberger and Peter(2017)]{Brandenberger}Brandenberger, R., and Peter, P. (2017). Bouncing cosmologies: progress and problems. \textit{Foundations of Physics, 47} (6), 797-850. https://doi.org/10.1007/s10701-016-0057-0

\bibitem[Brisson and Ofman(2021)]{Brisson}Brisson, L., Ofman, S. (2021). The two-triangle universe 
of Plato's timaeus and the in(de)finite diversity of the universe. \textit{Apeiron, 54} (4), 493-518.
https://doi.org/10.1515 \ /apeiron-2020-0064

\bibitem[Carnap(1947)]{Carnap}Carnap, R. (1947). \textit{Meaning and necessity}. Chicago: The University of 
Chicago Press.

\bibitem[Carone(2005)]{Carone}Carone, G. R. (2005). \textit{Plato's cosmology and its ethical dimensions}.
 Cambridge: Cambridge University Press.
 
\bibitem[Carroll(2020)]{Carroll} Carroll, J. W. (2020). Laws of nature. In  Edward N. Zalta (ed.), 
\textit{The Stanford Encyclopedia of Philosophy} (Winter 2020 Edition).
https://plato.stanford.edu/archives/win2020/entries \\ /laws-of-nature/. Accessed 2 Apr. 2022. 

\bibitem[Cartwright(1980)]{Cartwright}Cartwright, N. (1980). Do the laws of physics state the facts? 
\textit{Pacific Philosophical Quarterly, 61}, 75-84.
https://doi.org/10.1111/j.1468-0114.1980.tb00005.x

\bibitem[Dretske(1977)]{Dretske}Dretske, F. I. (1977). Laws of nature. \textit{Philosophy of Science, 44}, 248-268.

\bibitem[Einstein(1916)]{Einstein}Einstein, A. (1916). Die Grundlage der allgemeinen Relativitätstheorie. \textit{Annalen der Physik, 354} (7), 769-822. https://doi.org/10.1002/andp.19163540702

\bibitem[Eisenhauer(2005)]{Eisenhauer}Eisenhauer, F. et al. (2005). SINFONI in the galactic center: young stars and IR flares in the central light month. \textit{The Astrophysical Journal, 628}, 246-259.
https:// \ doi.org/10.1086/430667

\bibitem[Ellis et al.(2004)]{Ellis}Ellis, G. F. R., Kirchner, U., and Stoeger, W. R. (2004). 
Multiverses and physical cosmology. 
\textit{Monthly Notices of the Royal Astronomical Society, 347} (3), 921-936. https://doi.org/10.1111/j.1365-2966.2004.07261.x

\bibitem[Everett(1957)]{Everett}Everett, H. (1957). \enquote{Relative state} formulation of  
quantum mechanics. \textit{Reviews of Modern Physics 29}, 454. https://doi.org/10.1103/RevModPhys.29.454

\bibitem[Friedmann(1922)]{Friedmann}Friedmann, A. (1922). Über die Krümmung des Raumes. \textit{Zeitschrift für Physik A, 10} (1) 377. https://doi.org/10.1007/BF01332580

\bibitem[Glashow(1961)]{Glashow}Glashow, S. L. (1961). Partial symmetries of weak interactions.
\textit{Nuclear Physics 22}, 579-588. 
 https://doi.org/10.1016/0029-5582(61)90469-2

\bibitem[Guth(1981)]{Guth}Guth, A. H. (1981). The inflationary universe: a possible solution to the horizon and flatness problems. \textit{Physical Review D, 23}, 347-356. https://doi.org/10.1103/PhysRevD.23.347

\bibitem[Hume(1888)]{Hume}Hume, D. (1888). \textit{A treatise of human nature}. Oxford: Clarendon Press. 

\bibitem[Ijjas et al.(2013)]{Steinhardt}Ijjas, A., Steinhardt, P. J., and Loeb, A. (2013). Inflationary paradigm in trouble after Planck 2013. \textit{Physics Letters B, 723}, 261. https://doi.org/10.1016/j.physletb.2013.05.023

\bibitem[Kant(1998)]{Kant}Kant, I. (1998). \textit{Critique of pure reason}, 
translated by Paul Guyer and Allen W. Wood. Cambridge: Cambridge University Press.

\bibitem[Lehners(2008)]{Lehners}Lehners, J. L. (2008). Ekpyrotic and cyclic cosmology. \textit{Physics  Reports 465}, 223-263. https://doi.org/doi:10.1016/j.physrep.2008.06.001

\bibitem[Lemaître(1931)]{Lemaitre}Lemaître, G. (1931). A homogeneous universe of constant mass and increasing radius accounting for the radial velocity of extra-galactic nebulæ. \textit{Monthly Notices of the Royal Astronomical Society, 91}, 483. https://doi.org/10.1093/mnras/91.5.483

\bibitem[Lewis(1983)]{Lewis}Lewis, D. (1983). New work for a theory of universals. \textit{Australasian Journal of Philosophy, 61} (4), 343-377. https://doi.org/10.1080/00048408312341131

\bibitem[Linde(1983)]{Linde1983}Linde, A. (1983). Chaotic inflation. \textit{Physics Letters B 129}, 177-181.
https://doi.org/ \ 10.1016/0370-2693(83)90837-7

\bibitem[Linde(2015)]{Linde2015}Linde, A. (2015). Inflationary cosmology after Planck 2013. In C. Deffayet, P. Peter, B. Wandelt, M. Zaldarriaga, and L. F. Cugliandolo (eds), \textit{Post-Planck Cosmology: Lecture Notes of the Les Houches Summer School: Volume 100, July 2013}. Oxford: Oxford University Press. 

\bibitem[Neves(2019)]{Neves1}Neves, J. C. S. (2019). Nietzsche for physicists. \textit{Philosophia Scienti{\ae}, 23} (1), 185-201. https://doi.org/10.4000/philosophiascientiae.1855

\bibitem[Neves(2020)]{Neves2}Neves, J. C. S. (2020). Proposal for a degree of scientificity in cosmology. \textit{Foundations of Science, 25} (3), 857-878. https://doi.org/10.1007/s10699-019-09620-9

\bibitem[Neves(2021a)]{Neves3}Neves, J. C. S. (2021a). Cosmologias de antípodas: Platão e Nietzsche. 
\textit{Veritas (Porto Alegre) 66} (1), e37956. https://doi.org/10.15448/1984-6746.2021.1.37956

\bibitem[Neves(2021b)]{Neves4}Neves, J. C. S. (2021b). \textit{Demiurgos: sobre a criação de mundos}. 
São Paulo: Editora Livraria da Física. 

\bibitem[Nietzsche(2001)]{Nietzsche}Nietzsche, F. (2001). \textit{Gay science}, translated by Josefine Nauckhoff. Cambridge: Cambridge University Press. 

\bibitem[Nietzsche(2003)]{Nietzsche2}Nietzsche, F. (2003). \textit{Writings from the late notebooks},
translated by Kate Sturge. Cambridge: Cambridge University Press.

\bibitem[Novello and Perez Bergliaffa(2008)]{Novello}Novello, M. and Perez Bergliaffa S. E. (2008). Bouncing cosmologies. \textit{Physics Reports, 463}, 127-213. 
https://doi.org/10.1016/j.physrep.2008.04.006

\bibitem[Perlmutter et al.(1999)]{Perlmutter}Perlmutter, S. Supernova Cosmology Project et al. (1999).
Measurements of $\Omega$ and $\Lambda$ from 42 high redshift supernovae,
\textit{The Astrophysical Journal 517}, 565-586. 
https://doi.org/ \ 10.1086/307221

\bibitem[Planck(1900)]{Planck}Planck, M. (1900). Zur Theorie des Gesetzes der Energieverteilung im Normalspektrum.
\textit{Verhandlungen der Deutschen Physikalischen Gesellschaft 2}, 237-245.

\bibitem[Plato(1931)]{Plato}Plato. (1931). \textit{Timaeus}, translated by B. Jowett. London: Oxford University Press.

\bibitem[Popper(2005)]{Popper}Popper, K. (2005). \textit{The logic of scientific discovery}. London-New York: Routledge Classics.

\bibitem[Riess et al.(1998)]{Riess}Riess, A. G., Supernova Search Team et al. (1998)
Observational evidence from supernovae for an accelerating universe and a cosmological constant.
\textit{The Astronomical Journal 116}, 1009-1038. 
https://doi.org/10.1086/300499


\bibitem[Robertson(1935)]{Robertson}Robertson, H. P. (1935). Kinematics and world structure. \textit{Astrophysical Journal, 82}, 284. https://doi.org/10.1086/143681

\bibitem[Romero(2013)]{Romero}Romero, G. E. (2013). Adversus singularitates: the ontology of space-time singularities. \textit{Foundations of Science, 18} (2), 297-306. https://doi.org/10.1007/s10699-012-9309-4

\bibitem[Salam(1968)]{Salam}Salam, A. (1968). Weak and electromagnetic interactions. In N. Svartholm (Ed.),
\textit{Elementary particle theory: Relativistic groups and analyticity.
Proceedings of the eighth Nobel symposium} (pp. 367-377). Stockholm: Almquist \& Wiksell.

\bibitem[Starobinsky(1980)]{Starobinsky}Starobinsky, A. A. (1980). A new type of isotropic cosmological models without singularity. \textit{Physics Letters B, 91}, 99-102. https://doi.org/10.1016/0370-2693(80)90670-X

\bibitem[Swartz(1985)]{Swartz}Swartz, N. (1985). \textit{The concept of physical law}. Cambridge: Cambridge University Press. 

\bibitem[Tegmark(2004)]{Tegmark}Tegmark, M. (2004). Parallel universes. In J. Barrow, P. Davies, 
and C. Harper, Jr (Eds.), \textit{Science and ultimate reality: quantum theory, cosmology, and complexity} 
(pp. 459-491). Cambridge: Cambridge University Press.

\bibitem[Walker(1937)]{Walker}Walker, A. G. (1937). On Milne's theory of world-structure. \textit{Proceedings of the London Mathematical Society, 2-42} (1), 90. https://doi.org/10.1112/plms/s2-42.1.90

\bibitem[Weinberg(1967)]{Weinberg1}Weinberg, S. (1967). A model of leptons. \textit{Physical Review Letters, 19}, 1264-1266. https:// \ doi.org/10.1103/PhysRevLett.19.1264

\bibitem[Weinberg(2014)]{Weinberg2}Weinberg, S. (2014). \textit{Cosmology}. Oxford: Oxford University Press. 

\bibitem[Zeyl and Sattler(2019)]{Zeyl}Zeyl, D. and Sattler, B. (2019). Plato’s Timaeus. 
In  E. N. Zalta (ed.), \textit{The Stanford Encyclopedia of Philosophy} (Summer 2019 Edition).
https://plato.stanford.edu/archives/sum \\ 2019/entries/plato-timaeus/. Accessed 2 Apr. 2022. 


\end{thebibliography}
\end{document}